\documentclass[pre,final,onecolumn,showpacs,showkeys, preprintnumbers,amsmath,amssymb]{revtex4-2}

\usepackage{epsfig}
\usepackage{graphicx}
\usepackage{dcolumn}
\usepackage{graphics}
\usepackage{bm}
\usepackage{subfigure}
\usepackage{color}
\usepackage[normalem]{ulem}
\usepackage{soul}

\usepackage[utf8]{inputenc}

\begin{document}

\preprint{APS/123-QED}

\title{Time-fractional kinetic equation for the non-Markovian kinetic processes}

\author{E. Aydiner}
\email{ekrem.aydiner@istanbul.edu.tr}

\affiliation{Department of Physics, Faculty of Science, İstanbul University 34134, Istanbul Turkey}

\date{12 February 2021}

\begin{abstract}
In this study, we analytically formulated the path integral representation of the conditional probabilities for non-Markovian kinetic processes in terms of the free energy of the thermodynamic system. We carry out analytically the time-fractional kinetic equations for these processes. Thus, in a simple way, we generalize path integral solutions of the Markovian to the non-Markovian cases. We conclude that these pedagogical results can be applied to some physical problems such as the deformed ion channels, internet networks and non-equilibrium phase transition problems.
\end{abstract}


\maketitle


\section{Introduction}

Kinetic behaviour of the non-equilibrium, univariant or non-uniform systems is a very important problem in physics. In the literature, kinetic equations of these systems have been obtained by means of path integral formulations based on master equation \cite{Kubo1973,Kitahara1976a,Kitahara1976b,Kitahara1976c}. The non-equilibrium evolution of a thermodynamic system is presented by a Markovian master equation \cite{Lax1960,vanKampen1961},  
\begin{eqnarray} \label{eq:m-master}
	 \tau \frac{\partial P(x,t)}{\partial t} = - \sum_{\delta} W(x\rightarrow x+\delta x) P(x,t) + \sum_{\delta} W(x-\delta x\rightarrow x) P(x-\delta x,t)
\end{eqnarray}
where $P(x,t)$ is the finding probability at the $x$ point in the discrete space at the time $t$. The kernel $W(x\rightarrow x+\delta x)$ is the transition probability that represents changes in the discrete position variable  by an amount $\delta$ on the time scale $\tau$. In this theoretical framework, the free energy $F$ of the thermodynamic plays an important role in the dynamic of the system. Therefore, the kinetic transition probability is given depends on free energy by  
\begin{eqnarray} \label{eq:tr-prob}
	W(x\rightarrow x+\delta) = \exp(-\frac{\delta^{2}}{2\Delta}) \exp\{-\beta \frac{1}{2} [F(x+\delta) - F(x)] \}
\end{eqnarray}
where $\beta=(kT)^{-1}$ is the inverse temperature, $\Delta$ is a constant and $F$ is the free energy of the thermodynamic system. The kinetic transition probability in  Eq.~(\ref{eq:tr-prob}) firstly has been suggested by Langer \cite{Langer1969,Langer1971,Langer1973} as an extension of a model proposed by Glauber \cite{Glauber1963} based on Zwanzig theory \cite{Zwanzig1961,Zwanzig1964}.

The path integral solution of Eq.~(\ref{eq:m-master}) is given by \cite{Kubo1973,Kitahara1976a,Kitahara1976b,Kitahara1976c}
\begin{eqnarray} \label{eq:m-sol}
	P(x,t+\Delta t) = \int dx_{0}  K(x, t+\Delta t|x_{0},t) P(x_{0},t) + \mathcal{O}(\Delta t / \tau)^{2}
\end{eqnarray}
where the conditional probability $K(x,t|x_{0},t_{0})$ can be written as a path integral \cite{Kubo1973,Kitahara1976a,Kitahara1976b,Kitahara1976c},
\begin{eqnarray}
	K(x,t|x_{0},t_{0}) = \int \mathcal{D}(x) \exp \left[ - \int_{t_{0}}^{t} dt^{\prime} L(x(t^{\prime}), \dot{x}(t^{\prime})    \right]
\end{eqnarray}
where $\int \mathcal{D}(x)$ denotes the path integration and the argument of exponential term is action integral $L \left(x(t^{\prime}), \dot{x}(t^{\prime}) \right)$ is a Lagrangian of the kinetic motion which is a function of $x$ and $\dot{x}$. It is given in Refs.\,\cite{Kubo1973,Kitahara1976a,Kitahara1976b,Kitahara1976c}.
\begin{eqnarray}\label{eq:m-Ross}
	L(x,\dot{x}) =  
	\dot{x} (t) -  \Gamma \beta \frac{\partial F(x(t))}{\partial x}  = 0
\end{eqnarray}
or in Ref.\,\cite{Kubo1973} is given by
\begin{eqnarray}\label{eq:m-Kubo}
	L(x,\dot{x}) =  
	\left[ \dot{x} (t) -  \Gamma \beta \frac{\partial F(x(t))}{\partial x}   \right]^{2} - \Gamma^{\prime} \mu  \beta \frac{\partial^{2} F}{\partial x^{2}} = 0
\end{eqnarray}
where $\mu$ is an arbitrary parameter. As can be seen from Eqs.\,(\ref{eq:m-Ross}) and (\ref{eq:m-Kubo}) kinetic equations are clearly represented depend on the free energy of the system. Path integral formulation of the Markovian kinetic processes for the non-equilibrium system allows us to write the kinetic equations in terms of the free energy. 

Path integral formulation of conditional probability for the Markovian master equation have been extensively studied in literature \cite{Lax1960,vanKampen1961,Langer1969,Langer1971,Langer1973,Glauber1963,Zwanzig1961,Zwanzig1964,Kubo1973,Kitahara1976a,Kitahara1976b,Kitahara1976c,Graham1976,Weber2017}. By using this method, many differential equations have been solved. These discussions are given in a perfect review paper \cite{Weber2017}. On the other hand, path integral solutions for non-Markovian processes have been discussed in some references \cite{Pesquera1983,Hanggi1989,McKane1990,Bray1990,Luckock1990}.
However, so far, according to our knowledge, path integral formulation of the non-Markovian kinetic process depend on the free energy of the system has never been discussed. 

In this study, following the method presented in Refs.\,\cite{Kitahara1976a,Kitahara1976b,Kitahara1976c,Kubo1973} we generalize path integral solutions of Markovian to non-Markovian cases. We introduce path integral representation of conditional probabilities for non-Markovian kinetic processes in terms of free energy of the system. We carry out time-fractional kinetic equations depend on free energy. Analytical results are given in Section II, and finally conclusions are given in the last section.

\section{Time-fractional kinetic solution for the non-Markovian process}

Now, we can discuss the kinetic behaviour for the non-Markovian process of non-equilibrium systems. To proceed with the discussion, firstly, we briefly give the information about the master equation for non-Markovian process. The probability density function of a test particle at position $x$ and at time $t$ can be presented in terms of a generalized master equation (GME) as \cite{Metzler2000}
\begin{eqnarray}\label{eq:nm-master}
	\frac{\partial P(x,t)}{\partial t } =  \int_{-\infty}^{\infty} dx^{\prime} \int_{0}^{t} dt^{\prime} \mathcal{W}(x|x^{\prime},t-t^{\prime}) P(x^{\prime},t^{\prime})
\end{eqnarray}
where the kernel $\mathcal{W}(x|x^{\prime},t-t^{\prime})$ is the transfer probability from the position $x^{\prime}$ to $x$. As seen from Eq.\,(\ref{eq:nm-master}) that GME is very different from Eq.Eq.\,(\ref{eq:m-master}). Here, we discuss the probability function of the generalized master equation kernel $\mathcal{W}(x|x^{\prime},t-t^{\prime})$ by reformulated in terms of the path integral formulation based on previous approximations \cite{Kubo1973,Kitahara1976a,Kitahara1976b,Kitahara1976c}.

It is well known that Eq.\,(\ref{eq:nm-master}) takes, in Fourier-Laplace space, on the form
\begin{eqnarray}\label{eq:nm-LF}
	uP(k,u) - W_{0} (k) = \mathcal{W}(k,u) * P(k,u)
\end{eqnarray}
where $u$ is the Laplace variable, $k$ is the wave number and $f(x)*g(x)\equiv \int_{-\infty}^{\infty} dx^{\prime}f(x-x^{\prime})g(x^{\prime})$
denotes a Fourier convolution of the $f$ and $g$ functions. Dividing by $u$, after Laplace inversion and differentiation $\frac{\partial}{\partial t}$ we obtain another representation
\begin{eqnarray}\label{eq:nm-masterI}
	\frac{\partial P(x,t)}{\partial t } =  \frac{\partial}{\partial t} \int_{-\infty}^{\infty} dx^{\prime} \int_{0}^{t} dt^{\prime} \mathcal{W}(x|x^{\prime},t-t^{\prime}) P(x^{\prime},t^{\prime})
\end{eqnarray}
of the Eq.\,(\ref{eq:nm-master}). 

For Markovian processes transition probability $W(x\rightarrow x^{\prime})$ in Eq.~(\ref{eq:tr-prob}) is given in Gaussian form. However, in the presence of memory effect, the jump length $w(x)$ and waiting time $\Pi(t)$ probability distribution must be defined separately in the transition probability function $W(x\rightarrow x^{\prime})$. Let us definition of the transition probability for non-Markovian process as $\mathcal{W}(x|x^{\prime},t-t^{\prime}) = \Pi(t) \ w(x)$ which cover two different dynamics where the transfer kernel $w(x|x^{\prime})$ is denotes for spatial correlations and the memory kernel introduces the non-Markovian dynamics, independently. For example, in a such dynamical process, the waiting time distribution leads to time-fractional form. However, jump length distribution produce Levy flight leads to space-fractional dynamics \cite{Metzler2000}. In this section, we study path integral solution for the different dynamics and obtain time and space-fractional kinetic equations depend on the free energy of the system.

The transition probability $\mathcal{W}$ is defined as $\mathcal{W}(x\rightarrow x+\delta x,t-t^{\prime}) = w(x|x^{\prime})\Pi(t)$ mentioned above. The waiting time distribution $\Pi(t)$ is given by
\begin{eqnarray}\label{eq:nm-powerd}
	\Pi (t) = \frac{1}{\Gamma(\gamma)} \left(\frac{t}{\tau}\right)^{\gamma-1}
\end{eqnarray}
where $0<\gamma<1$ for the anomalous sub-diffusion. In this case, the new kernel is given by 
\begin{eqnarray}\label{eq:nm-trans}
	\mathcal{W}(x\rightarrow x+\delta x,t-t^{\prime}) = \frac{1}{\Gamma(\gamma)} \left(\frac{t}{\tau}\right)^{\gamma-1} w(x|x^{\prime})
\end{eqnarray}
where $w(x|x^{\prime})$ is responsible for spatial correlations, which is defined as
\begin{eqnarray} \label{strans}
	w(x|x^{\prime})= \exp(-\frac{\delta^{2}}{\Delta}) \exp\{-\beta \frac{1}{2} [F(x+\delta) - F(x)]
\end{eqnarray}
It is known that the $w(x|x^{\prime})$ and $\Pi(t)$ are independent. The solution of the GME shows a strong dependence features a strong dependence on its stochastic history. The resulting equation is
\begin{eqnarray}\label{eq:nm-result}
	\frac{\partial P(x,t)}{\partial t } =  \frac{1}{\Gamma(\gamma)} \frac{\partial}{\partial t} \int_{0}^{t} dt^{\prime} \left(t-t^{\prime}\right)^{\gamma-1} \int_{-\infty}^{\infty} dx^{\prime}  w(x|x^{\prime}) P(x^{\prime},t^{\prime})
\end{eqnarray}
Eq.~(\ref{eq:nm-result}) includes the defining expression \cite{Oldham1974}
\begin{eqnarray}\label{eq:nm-fract}
	_{0}D^{1-\gamma}_{t} P(x,t) = \frac{1}{\Gamma(\gamma)} \frac{\partial}{\partial t} \int_{0}^{t} dt^{\prime}  \left(t-t^{\prime}\right)^{\gamma-1} P(x,t^{\prime})
\end{eqnarray}
where $_{0}D^{1-\gamma}_{t}$ is the Rieman-Liouville fractional derivative. Time-fractional master equation can be expressed in te form 
\begin{eqnarray}\label{eq:nm-timef}
	\frac{\partial P(x,t)}{\partial t } =   _{0}D^{1-\gamma}_{t}  \int_{-\infty}^{\infty} dx^{\prime}  w(x|x^{\prime}) P(x^{\prime},t^{\prime}) \ .
\end{eqnarray}
The discrete form of Eq.~(\ref{eq:nm-timef}) is given by
\begin{eqnarray}\label{eq:nm-tfdiscrete}
	\frac{\partial P(x,t)}{\partial t } =   _{0}D^{1-\gamma}_{t}  \sum_{x^{\prime}}  w(x|x^{\prime}) P(x,t) \ .
\end{eqnarray} 
To proceed calculation we write Eq.~(\ref{eq:nm-tfdiscrete}) as
\begin{eqnarray} \label{Master}
	\frac{\partial P(x,t)}{\partial t } = - _{0}D^{1-\gamma}_{t} H(x,\partial_{x})P(x,t)
\end{eqnarray}
where $h(x,\partial_{x})$
\begin{eqnarray} \label{eq:nm-with}
	H(x,\partial_{x}) P(x,t) = \sum_{\delta} (1 - e^{-\delta \partial/\partial x} ) w (x\rightarrow x+\delta) P(x,t)
\end{eqnarray}
this relation causes the Kramers-Moyal expansion \cite{Kramers1940,Moyal1949}
\begin{eqnarray} \label{eq:tf-km}
	H(x,\partial_{x}) P(x,t)= \sum_{m=1}^{\infty} \frac{(-1)^{m+1}}{m !} \delta^{n} \frac{\partial^{m}}{\partial x^{m}} \sum_{\delta} \delta^{m} w (x\rightarrow x+\delta) P(x,t) \ .
\end{eqnarray}
where $\frac{\partial^{m}}{\partial x^{m}}$ operates at the same time both $w (x\rightarrow x+\delta) P(x,t)$ and $w (x\rightarrow x+\delta)$. The sums are
over all possible values of the multi-indices $m$. The derivative in the left side of Eq.(\ref{Master}) can be written as a short time solution. Therefore we can obtain $P(x,t+\Delta t)$ as
\begin{eqnarray} \label{new-P-H}
	P(x,t+\Delta t) =  \{ 1 + \Delta t \ _{0}D^{1-\gamma}_{t}   \sum_{m=1}^{\infty} \frac{(-1)^{m+1}}{m !} \delta^{n} \frac{\partial^{m}}{\partial x^{m}} \sum_{\delta} \delta^{m} w (x\rightarrow x+\delta \} P(x,t)  
\end{eqnarray}
We define the Fourier transform $\mathcal{F} \{P\}(k,t+\Delta t)$ of Eq.(\ref{new-P-H}) as
\begin{eqnarray}\label{Fourier}
	\mathcal{F} \{P\}(k,t+\Delta t) = \mathcal{F} \{P\}(k,t)  + \Delta t \ _{0}D^{1-\gamma}_{t} \sum_{m=1}^{\infty} \frac{(-1)^{m+1}}{m !} \delta^{m} k^{m} (-i)^{|m|}  \} \mathcal{F} \{wP\}(k,t)  
\end{eqnarray}
This of course can be written as
\begin{eqnarray}\label{F-trans}
	\mathcal{F} \{P\}(k,t+\Delta t) = (2\pi)^{-1/2}  \int_{-\infty}^{\infty} dx_{0} e^{ikx_{0}} \{ 1 + \Delta t \ _{0}D^{1-\gamma}_{t} H(x_{0},-ik) \} P(x_{0},t)  
\end{eqnarray}
where $h(x_{0},-ik)$ is obtained from Eq.\,(\ref{eq:tf-km}) by replacing $\partial/\partial x$ with $-ik$ and $x$ with $x_{0}$ \cite{Kitahara1976a,Kitahara1976b,Kitahara1976c}.
On the other hand, the inverse Fourier transform of Eq.\,(\ref{F-trans}) is given by
\begin{eqnarray} \label{eq:nm-f1}
		P(x,t+\Delta t)  =  (2\pi)^{-1/2} \int dk e^{-ikx} \mathcal{F} \{P\}(k,t+\Delta t) \ .
\end{eqnarray} 
Here, introducing Eq.\,(\ref{F-trans})) into Eq.\,(\ref{eq:nm-f1}) and recognizing that for small $\Delta t$ the curly bracket in Eq.\,(\ref{F-trans}) is an exponential, in this case, we obtain
\begin{eqnarray} \label{path}
	P(x,t+\Delta t) = (2\pi)^{-m} \int_{-\infty}^{\infty} dx_{0} \int_{-\infty}^{\infty} dk \exp \left[ -	\Delta t \{ ik(x-x_{0})/ \Delta t \} -  _{0}D^{1-\gamma}_{t} H(-ik, x_{0}) \right]  P(x_{0},t)
\end{eqnarray}
The kernel of Eq.\,(\ref{path}) can be defined as
\begin{eqnarray} \label{Kernel}
 K_{\gamma}(x, t|x_{0},t) =  \int  \mathcal{D} (x_{0}) \int  \mathcal{D}(k)   \exp \left[ - \int_{t_{0}}^{t} dt^{\prime} \{ ik(t^{\prime}) \dot{x} (t^{\prime}) \} -  _{0}D^{1-\gamma}_{t^{\prime}} H(-ik(t^{\prime}), x(t^{\prime}) \right]
\end{eqnarray}
where $\dot{x}(t^{\prime}) = (x-x_{0})/ \Delta t$. This kernel represents the path integral formulation of the conditional probability for non-Markovian kinetics. We clearly see that the integral argument in Eq.(\ref{Kernel}) corresponds to Lagrangian of the system, which is given as
\begin{eqnarray} \label{Lagrangian}
	L_{\gamma}(k, x, \dot{x}) =  ik(t^{\prime}) \dot{x} (t^{\prime}) \} -  _{0}D^{1-\gamma}_{t^{\prime}} H \left(-ik(t^{\prime}), x(t^{\prime}) \right) 
\end{eqnarray}
where $H \left(-ik(t^{\prime}), x(t^{\prime}) \right) $ can be read as Hamiltonian. 
The path integral  in Eq.(\ref{Kernel}) can be defined as the limit of the multiple
integral
\begin{eqnarray} \label{Multi}
\int dx_{L-1}... \int dx_{1} \int dk_{L-1} ... \int dk_{0} (2\pi)^{-mL} \exp \left[  -\Delta t \sum_{j=0}^{L-1} \{ \langle   i k^{j}, (x^{j+1}-x^{j} ) / \Delta t  \rangle  - _{0}D^{1-\gamma}_{t^{\prime}} H(-ik^{j},x^{j} )  \}     \right]
\end{eqnarray}
when $L=(t-t_{0}) / \Delta t \rightarrow \infty$. 
We consider here that the transition between the small paths along the trajectory are independent of each other. Now, by using Eq.(\ref{strans}) we can write $H(-ik,x)$ as
\begin{eqnarray} \label{Hk}
	H \left(-ik, x) \right)= i \sum_{j=1}^{m} k_{j} \int d\delta \delta_{j} \exp (-\Delta^{-1} \langle\delta,\delta \rangle) \exp\left[ -\beta \frac{1}{2}  \{ F(x+\delta) - F(n) \}  \right] 
	-  \frac{1}{2} \sum_{j=1}^{m} \sum_{i=1}^{m}  k_{j} k_{i} 
	\int d\delta \delta_{j} \delta_{i} \nonumber \\ 
	\exp (-\Delta^{-1} \langle\delta,\delta \rangle ) \exp\left[ -\beta \frac{1}{2}  \{ F(x+\delta) - F(n) \}  \right]
\end{eqnarray}
For very small $\Delta$ values, the integrations can be obtained as
\begin{eqnarray} \label{Hamiltonian}
	H \left(-ik(t^{\prime}), x(t^{\prime}) \right)= - \frac{1}{2} \beta \Gamma \sum_{j=1}^{m} i k_{j} \frac{\partial F}{\partial x_{j}} - \frac{1}{2} \Gamma \sum_{j=1}^{m} k^{2}_{j}
\end{eqnarray}
with 
\begin{eqnarray} \label{Gamma}
	\Gamma = \left( 2\pi \Delta \right)^{m/2} \Delta \ .
\end{eqnarray}
Introducing  Eq.\,(\ref{Hamiltonian}) into Eq.\,(\ref{Multi}) and integrating over $k$, we get the time-fractional kernel 
\begin{eqnarray}\label{frac-kernel}
	K_{\gamma}(x,t|x_{0},t_{0}) 
= \int \mathcal{D} (x) \exp \left[ - \int_{t_{0}}^{t}  dt^{\prime} 
\{ _{0}D^{\gamma}_{t} x (t^{\prime}) + \Gamma \beta \frac{\partial F(x(t^{\prime}))}{\partial x}   \}^{2} \right] \ .
\end{eqnarray}
The path of extreme probability are those for which $\int_{0}^{\tau}dt^{\prime} L_{\gamma}(x,\dot{x},k)$ is extremized, namely, time-fractional action integral satisfy the condition
\begin{eqnarray} \label{action}
	\delta \int_{t_{0}}^{t} dt^{\prime} L_{\gamma} [x(t^{\prime}),\dot{x}(t^{\prime})] = 0
\end{eqnarray}
where $L_{\gamma}(x,\dot{x})$ is the time-fractional Lagrangian of the system.  Euler–Lagrange equation in the action Eq.\,(\ref{action})) can be solved due to $k$ as a function of $x$ and $\dot{x}$ as
\begin{eqnarray} \label{E-L}
	\frac{\partial L_{\gamma}}{\partial k} = 0 \ .
\end{eqnarray}
The solution of the it gives the time-fractional kinetic equation
\begin{eqnarray}\label{eq:nm-m0}
	_{0}D^{\gamma}_{t} x (t) +  \Gamma \beta \frac{\partial F(x(t))}{\partial x}  = 0  
\end{eqnarray}
where $0<\gamma<1$. As can be seen from Eq.(\ref{eq:nm-m0}) that the time evolution of the stochastic variable $x$ is governed by integro-differential operator. One can easily see that the time dependent dynamics of the system deviates from exponential. The operator $_{0}D^{\gamma}_{t}$ points out that the solution of the kinetic equation Eq.\,(\ref{eq:nm-m0})) possesses Mittag-Leffler function.

\section{Conclusion}

In this study, considering the non-Markovian kinetic processes, we presented the path integral formulation of the conditional probability depends on the free energy of the system. Following the method given in Refs.\cite{Zwanzig1961,Zwanzig1964,Langer1969,Langer1971,Langer1973,Kubo1973,Kitahara1976a,Kitahara1976b,Kitahara1976c}, we show that path integral solutions of the Markovian can easily be generalized to the non-Markovian cases for kinetic processes. We analytically obtain the time-fractional kinetic equations for these processes in the case of memory effects. We conclude that the time-fractional kinetic equation can be applied to the some realistic physical problems to analyse the dynamics of them depend on the free energy such as the deformed ion channels, internet networks and non-equilibrium phase transition problems.

\section{Acknowledgement}

A part of this work had been completed during visiting Potsdam University. Therefore, I would like to thank Professor Ralf Metzler and Potsdam University for their kind hospitality.

\bibliography{path}

\end{document}